\documentclass[11pt]{article}
\usepackage{amsmath}
\usepackage{amssymb}
\usepackage{graphics}
\begin{document}

\title{Thermodynamics of Black Holes in Schr\"odinger Space}
\author{D.~Yamada
        \bigskip
        \\
       {\it Racah Institute of Physics},
       {\it The Hebrew University of Jerusalem},
        \smallskip
       \\
         {\it Givat Ram, Jerusalem, 91904 Israel}
          \bigskip
       \\
         {\tt daisuke@phys.huji.ac.il}}
\date{}
\maketitle

\begin{abstract}
A black hole and a black hyperboloid
solutions in the space with the
Schr\"odinger isometries are presented
and their thermodynamics is examined.
The on-shell action is obtained by the difference between the extremal and
non-extremal ones with the unusual matching of the boundary metrics.
This regularization method is first applied to
the black brane solution in the space of the Schr\"odinger symmetry 
and shown to correctly reproduce the known thermodynamics.
The actions of the black solutions
all turn out to be the same as the AdS counterparts.
The phase diagram of the black hole
system is obtained in the parameter space
of the temperature and chemical potential and
the diagram contains the Hawking-Page phase transition and instability lines.
\end{abstract}

\bigskip

\tableofcontents
\pagebreak

\section{Introduction}

Pioneered by Son~\cite{Son:2008ye}, and
Balasubramanian and McGreevy~\cite{Balasubramanian:2008dm},
the developments in the Galilean holography have been seen lately.
They are aimed to be
the non-relativistic generalizations of the AdS/CFT correspondence
\cite{Maldacena:1997re,Gubser:1998bc,Witten:1998qj}.

The generators of the Galilean algebra correspond to the temporal and
spatial translations, rotational transformations, as well as to the Galilean
boost and mass operators.
One can extend the algebra by including the dilatation operator,
and for the special case where the powers of the time coordinate
scale twice as much as the spatial coordinates, the full non-relativistic
conformal symmetry, that includes the special conformal transformation,
can be obtained.
The non-relativistic conformal symmetry is also known as the
Schr\"odinger algebra \cite{Hagen:1972pd,Niederer:1972zz}.%
\footnote{
  Also see Reference~\cite{Nishida:2007pj} and the references
  therein.
}
The authors of References~\cite{Son:2008ye,Balasubramanian:2008dm}
have discovered the $d+3$ dimensional spacetime
geometry whose isometry group is the Schr\"odinger symmetry
of $d$ spatial dimensions, and
based on the AdS/CFT correspondence,
they proposed that the gravitational system of this 
Schr\"odinger geometry
is the holographic dual of the non-relativistic conformal
field theories at strong couplings.
The expectation is that the duality would provide the important
information about the strongly coupled real world
non-relativistic systems (see Reference~\cite{Son:2008ye}
for the examples of such systems), as it has been the case
for the AdS/CFT correspondence.

Among the subsequent developments, the finite temperature
generalization of the Galilean holography was discussed in
References~\cite{Herzog:2008wg,Maldacena:2008wh,Adams:2008wt}.
In parallel to the finite temperature
AdS/CFT correspondence \cite{Witten:1998zw},
they have proposed a planar black solution, a black brane,
in the Schr\"odinger space as the holographic dual of the
non-relativistic CFT at finite temperature.
The authors embed the holographic set up in string theory.
References~\cite{Herzog:2008wg,Adams:2008wt} 
start with the near horizon geometry of D3-branes
which is the AdS$_5$ black brane times $S^5$.
Then they
apply a solution generating procedure known as the Null Melvin
Twist \cite{Alishahiha:2003ru,Gimon:2003xk} to this system,
and upon the KK-reduction of $S^5$,
they find that the resulting
geometry is a black brane solution whose extremal and
asymptotic limits reduce to the Schr\"odinger geometry.%
\footnote{
  One notices in References~\cite{Herzog:2008wg,Adams:2008wt} that
  the dynamical exponent $\nu$ of the resulting geometry is $1$.
  The dynamical exponent $\nu$ determines the relative scaling of
  the time coordinate and others.
  For $\nu=1$, the powers of the scaling for the time is twice
  that of translationally invariant coordinates and as noted,
  this is the special case where
  the Galilean symmetry extends to the Schr\"odinger.
}
Notice that by construction, this procedure yields the planar
black solutions.
The thermodynamic and transport properties were discussed
in the references and References~\cite{Herzog:2008wg,Maldacena:2008wh}
noticed the similarity to the usual Schwarzschild-AdS black branes.

Just as for the asymptotically AdS case, the planar black solution
does not exhibit the Hawking-Page phase transition
\cite{Hawking:1982dh}.
The phase transition, in the AdS case, is observed for 
the black holes, as opposed to the black branes,
and the phase structure of the black holes tends to be richer than
the planar solutions.
Moreover,
in the context of the AdS/CFT correspondence,
the dual field theory is supposed to go through
the confinement/deconfinement phase transition, corresponding
to the gravitational Hawking-Page phase transition
\cite{Witten:1998zw}.
Therefore, it is natural to search for the black hole solution
in the Schr\"odinger space and this is the topic of this paper.

Our approach to this problem is not to embed the story in string
theory but to directly deal with the $(d+3)$-dimensional action
whose solution is supposed to be dual to the $d$ spatial
dimensional non-relativistic conformal field theory.
(To be concrete, we will concentrate on the case with $d=2$.)
Such an action has been proposed by Son~\cite{Son:2008ye} and
by the authors of
References~\cite{Herzog:2008wg,Maldacena:2008wh,Adams:2008wt}.
We seek for the black hole solution to this action and we indeed
find one by mimicking the relationship between the black brane
and black hole in AdS.

To examine the thermodynamics of a given black solution, we need to
obtain the value of the action and the thermodynamic quantities.
Because of the asymptotic structure of the Schr\"odinger space
and the volume integral which must be carried out,
the on-shell value of the action diverges
and we must somehow regularize this.
For this purpose,
Reference~\cite{Herzog:2008wg} proposed boundary terms,
similarly to the counterterm technique of asymptotically AdS systems.
This proposal has a few problems.
One problem is that the resulting finite action seems
to depend on arbitrary coefficients of the boundary terms introduced.
In the reference, some minimal number of the boundary terms were
considered, however, there are more terms that can contribute to
the finite action [{\it e.g.}, $(A_\mu A^\mu)^2$].
This ambiguity may be eliminated by requiring the first law to be
satisfied for the resulting thermodynamic quantities.
However, given the action with the counterterms, it is natural to
compute the thermodynamic quantities through the definition
proposed by Brown and York \cite{Brown:1992br}.
(Reference~\cite{Herzog:2008wg} computes
thermodynamic quantities differently,
as we will do in Section~\ref{subsec:BBThermo}.)
This attempt fails unless one chooses unjustifiably awkward
normalization for the time-like boundary Killing vector field.
The most important problem for this paper is that the similar
counterterm technique miserably fails for our black hole solution.
The counterterms in this case are not in the form to cancel the
divergences coming from the bulk action and the Gibbons-Hawking
surface term, mainly because of the nonzero scalar field
(dilaton) at infinity.

Another way to obtain the thermodynamic quantities was proposed
in Reference~\cite{Maldacena:2008wh}.
The idea was to transform the metric to the asymptotically AdS
form by using the symmetry of the system.
(The symmetry actually is that of the eight dimensional supergravity.
See the appendix of the reference for the details.)
Then compute the relevant quantities by utilizing the established
method \cite{Balasubramanian:1999re,Kraus:1999di,deHaro:2000xn,Bianchi:2001kw}
for the AdS case.
This procedure, however, does not directly apply for the black hole solution,
mainly because the horizon of the black hole
does not possess the spatial translation invariance.
(However, see the discussion in Section~\ref{sec:discussions} where
we argue that there should be a similar procedure that is applicable
for the black hole.)

We therefore propose a new regularization method that is apt
for all the black solutions discussed in this paper.
This actually is the oldest regularization technique
proposed by Gibbons, Hawking and
Page \cite{Gibbons:1976ue,Hawking:1982dh}, namely,
the subtraction method,
but with an unusual boundary matching.
This method proceeds as follows.
One needs to evaluate the action (with Gibbons-Hawking surface term)
on the spaces with and without
the black hole but due to the volume integral,
each action is divergent.
To regulate the divergences, one put the systems
in a large box and the metrics on
the wall of the box (the first fundamental forms) must be matched.
Then one subtracts the regulated action from the other
and the wall of the box is removed to infinity.
In the Schwarzschild-AdS black hole case
the difference is finite and
the sign of the difference action determines the thermodynamically
preferred solution (that is, it is suitable in searching
for the Hawking-Page phase transition).
In our geometry, we find a problem for this procedure:
the boundary matching cannot be achieved because of the peculiar
nature of one of the coordinates.
Therefore, in this paper,
we propose a kind of ``partial matching'' of the first fundamental forms
where we scale the coordinates on the wall to match the metrics,
except for the special coordinate.

In Section~\ref{sec:BB},
the procedure is first applied to the known black brane case and
checked to reproduce the thermodynamics of
References~\cite{Herzog:2008wg,Maldacena:2008wh}.
It is noteworthy that Reference~\cite{Herzog:2008wg} and we find
the finite action be the same as that of the
Schwarzschild-AdS black brane.
Next in Section~\ref{sec:BH}, the black hole solution is presented
and the same procedure is applied to this solution.
Remarkably, the resulting action turns out to be the same as
the one for the Schwarzschild-AdS black hole.
The thermodynamic quantities resulting from the action are
computed and the phase diagram of the system is obtained.
Among other structures in the phase diagram, we find the Hawking-Page
phase transition.
Section~\ref{sec:discussions} is dedicated to the discussions.
The appendix presents the black hyperboloid solution and
examines its thermodynamic properties.

\section{Black Brane}\label{sec:BB}
In this section, we mainly review the results obtained in
References~\cite{Herzog:2008wg,Maldacena:2008wh}.
We, however, propose a new way of computing the finite on-shell action
in Section~\ref{subsec:BBDiffAct}.
The same method will be applied to the black hole solution in
the next section and this section serves to show the validity
of the method.

\subsection{The Solution}\label{subsec:BBSoln}

In References~\cite{Herzog:2008wg,Maldacena:2008wh,Adams:2008wt},
the Schr\"odinger geometry discovered in
References~\cite{Son:2008ye,Balasubramanian:2008dm}
was embedded in the frame work of string theory.
References~\cite{Herzog:2008wg,Adams:2008wt} achieve this by
applying the Null Melvin Twist procedure \cite{Alishahiha:2003ru,Gimon:2003xk}
to the D3-brane geometry.
Let us start with the non-extremal D3-brane geometry in the near horizon region
(Schwarzschild-AdS black brane):
\begin{align}\label{eq:1stMetric}
  ds^2 &= \bigg(\frac{r}{R}\bigg)^2 \big(-h\,dt^2+dy^2+d\vec{x}^2\big)
         +\bigg(\frac{R}{r}\bigg)^2 h^{-1}dr^2 + R^2d\Omega_5^2
  \;,\nonumber\\
  \Phi &= 0
  \;,\nonumber\\
  B &= 0
  \;,
\end{align}
where $R$ is the AdS scale, $\vec{x}:=(x_1,x_2)$, $h:=1-(r_H/r)^4$,
and $d\Omega_5^2$ is the line element of the unit five sphere.
The location $r=r_H$ is where the horizon is and notice that
setting $r_H=0$ reduces the metric to the extremal case.
The fields $\Phi$ and $B$ are the dilaton and NSNS two-form,
respectively.
We have the RR four-form potential as well but it does not play
an important role in our discussion and hence omitted.
For convenience, References~\cite{Herzog:2008wg,Maldacena:2008wh,Adams:2008wt}
adopt the $U(1)$ Hopf fibration over
$\mathbb{P}^2$ to write
\begin{equation}
  d\Omega_5^2 = \eta^2 + ds_{\mathbb{P}^2}^2
  \;,
\end{equation}
where $\eta$ is the one-form in the Hopf fiber direction and
$ds_{\mathbb{P}^2}^2$ is the metric on $\mathbb{P}^2$
(see Appendix A of Reference~\cite{Adams:2008wt} for a little more details).
As always, $\eta^2$ is understood to be the symmetric tensor
product.
(When we intend to mean alternating projections, we write
wedges explicitly.)

The coordinate $y$ is arbitrarily singled out from the three
translationally invariant spatial directions
(the other directions are denoted by $\vec{x}$) and it is the direction
that plays a special role in the Null Melvin Twist procedure.
The procedure yields the system,
\begin{align}\label{eq:k0tybasis}
  ds^2 =& K^{-1}\bigg(\frac{r}{R}\bigg)^2 
          \bigg[-(1+b^2r^2)\,h\, dt^2 -2b^2r^2 h\, dt dy
          +(1-b^2r^2h) dy^2 + K d\vec{x}^2 \bigg]
          \nonumber\\
          &+ \bigg(\frac{R}{r}\bigg)^2 h^{-1}dr^2 
          + K^{-1}R^2\eta^2 + R^2ds_{\mathbb{P}^2}^2
  \;,\nonumber\\
  \Phi =& -\frac{1}{2}\ln K
  \;,\nonumber\\
  B =& K^{-1}\bigg(\frac{r}{R}\bigg)^2 b\, (\,h\, dt + dy)\wedge\eta
  \;,
\end{align}
where $K:=1-(h-1)b^2r^2$
and the parameter $b$ has the dimension
of inverse length and is related to
the twist factor given in $\eta$-direction (see
References~\cite{Alishahiha:2003ru,Gimon:2003xk,Herzog:2008wg,Adams:2008wt}).
The KK reduction of the five-sphere is a consistent
truncation \cite{Maldacena:2008wh} and this
provides a five dimensional system.
In doing so, following Herzog {\it et al.} \cite{Herzog:2008wg},
we introduce the light-cone coordinates
\begin{equation}\label{eq:LC}
  x^+ := bR(t+y) \;, \quad \text{and} \quad
  x^- := \frac{1}{2bR} (t-y)
  \;.
\end{equation}
The particular choice of the normalization will be explained shortly.
We then have the five dimensional system
\begin{align}\label{eq:k0nonExt}
  ds^2 =& K^{-2/3} \bigg(\frac{r}{R}\bigg)^2
          \bigg[ -\bigg\{\frac{h-1}{(2bR)^2}
          + \bigg(\frac{r}{R}\bigg)^2 h \bigg\} dx^{+2}
          - (1+h) dx^+dx^-
          + (bR)^2(1-h)dx^{-2}
        \nonumber\\
        &+ K d\vec{x}^2 \bigg]+ K^{1/3} \bigg(\frac{R}{r}\bigg)^2 h^{-1}dr^2
  \;,\nonumber\\
  \Phi =& -\frac{1}{2}\ln K
  \;,\nonumber\\
  A =& K^{-1}\bigg(\frac{r}{R}\bigg)^2 b \,
     \bigg\{ \frac{h+1}{2bR} dx^+ + bR(h-1) dx^- \bigg\}
  \;,
\end{align}
where we changed to the Einstein frame and renamed the KK-reduced
two-form field $B$ to $A$ which is a one-form field.
[The reader is warned for the slightly confusing notations here:
$dx^{+2}=(dx^+)^2$ and $dx^{-2}=(dx^-)^2$.]
As suggested by Son~\cite{Son:2008ye}, $x^+$ is reinterpreted
as the time coordinate and $x^-$ is assumed to be a compactified
direction to yield discrete mass spectrum of the system.

The extremal case is given by the value $h=1$ and the non-extremal
case approaches this at asymptotically large $r$.
Notice that the normalization of the light-cone coordinates in
Equations (\ref{eq:LC}) is designed to yield the extremal system
which is free of the parameter $b$.
This means that the parameter is not physical in the extremal case
and this fact makes the normalization of the light-cone coordinates
(\ref{eq:LC}) unique, even for the non-extremal case because
that should be fixed by the boundary metric.%
\footnote{
  This fact was pointed out to the author by Ofer Aharony and
  Zohar Komargodski.
}
\\

The five dimensional system (\ref{eq:k0nonExt}) can be obtained
from the equations of motion that follow from a five dimensional
action.
Such an action was originally proposed by Son~\cite{Son:2008ye}
for the extremal system and
References~\cite{Herzog:2008wg,Maldacena:2008wh,Adams:2008wt}
find the action that also supports the non-extremal system:
\begin{equation}\label{eq:theAction}
  S_5 = \frac{1}{16\pi G_5} \int dx^5 \sqrt{-g}
        \bigg[ \mathcal{R} 
             - \frac{4}{3}(\partial_\mu \Phi)(\partial^\mu \Phi)
             - \frac{1}{4} R^2 e^{-8\Phi/3}F_{\mu\nu}F^{\mu\nu}
             - 4 A_\mu A^\mu - V/R^2
        \bigg]
        \;,
\end{equation}
where $G_5$ is the five dimensional Newton constant, $g$ is the determinant
of the five dimensional metric, $\mathcal{R}$
is the scalar curvature, $F:=dA$ and
the potential $V$ is defined to be
\begin{equation}
  V := 4 e^{2\Phi/3}(e^{2\Phi}-4)
  \;.
\end{equation}
Son's original action is recovered by setting $\Phi$-field to zero.
\\

We wrap up this subsection by briefly discussing the geometric properties
of the metric, especially the causal development of it.
Because of the nonzero $g_{+-}$ component, the geometry is stationary
but not static.
We interpret this as the rotating black brane in the compactified
$x^-$ direction.
Let us rewrite the metric in the ADM form;
\begin{align}\label{eq:ADMform}
  ds^2 =& K^{-2/3} \bigg(\frac{r}{R}\bigg)^2
        \bigg[ - \bigg\{ \frac{1}{(bR)^2(1-h)} 
                        + \bigg(\frac{r}{R}\bigg)^2
                 \bigg\} h\, dx^{+2}
  \nonumber\\
                 &+ (bR)^2(1-h) \bigg\{
                                dx^- - \frac{1+h}{2(bR)^2(1-h)}dx^+
                              \bigg\}^2
                 + K d\vec{x}^2
        \bigg]
        + K^{1/3} \bigg(\frac{R}{r}\bigg)^2 h^{-1} dr^2
  \;.
\end{align}
From this form, we can pick up some information about this geometry.
First, recall that $h(r_H)=0$, so we clearly see the existence
of the ergo-region, as typical for a rotating black hole.
Second, the angular velocity of the horizon $\Omega_H$
measured in the units of $x^-$-circumference is given by
\begin{equation}\label{eq:OmegaH}
  \Omega_H = \frac{1}{2(bR)^2}
  \;,
\end{equation}
and the coordinate angular velocity diverges at the boundary
where $h\to 1$.

In the usual notion of geometrodynamics, $x^+$ is supposed to describe
the Cauchy development of the space-like four-surface with
the metric
\begin{equation}\label{eq:4surface}
  ds_4^2 = K^{-2/3}(br)^2(1-h) dx^{-2}
         + K^{1/3} \bigg(\frac{r}{R}\bigg)^2 d\vec{x}^2
         + K^{1/3} \bigg(\frac{R}{r}\bigg)^2 h^{-1} dr^2
  \;.
\end{equation}
This poses a problem for the extremal case with $h=1$,
because the $g_{4--}$ component is degenerate.
The surface of constant $x^+$ is not space-like
and the four-surface has zero measure.
This fact stems from the light-like nature of the coordinate $x^-$
before we reinterpret $x^+$ as the time coordinate.
Therefore, we see that $x^+$ is not an appropriate time function
in the five dimensional spacetime point of view.
One can, actually, see that this is not just due to the particular choice
of the time coordinate.
For example, if we use time $t$ as in (\ref{eq:k0tybasis}),
we see that the lapse function squared always becomes negative
at sufficiently large $r$.
The fact is that this spacetime is not globally hyperbolic,
just like AdS and pp-wave spacetimes and
to the latter, the Schr\"odinger geometry is conformal.
The causal pathology of the geometry (\ref{eq:k0nonExt}) was
pointed out in Reference~\cite{Herzog:2008wg} and argued to be
the evidence of the holographic dual of non-relativistic CFT
which should allow the action-at-a-distance.
Though we feel that this interesting point should be investigated
further, we will not dwell on this issue in this paper and leave
it for the future consideration.

\subsection{The Difference Action}\label{subsec:BBDiffAct}

As noted before, the black solution (\ref{eq:k0nonExt}) found
in References~\cite{Herzog:2008wg,Maldacena:2008wh,Adams:2008wt}
approaches
the extremal solution near the boundary and therefore naturally
interpreted as the finite temperature generalization of the Galilean
holography.
This is similar to the finite temperature AdS/CFT
correspondence \cite{Witten:1998zw}.
We thus would like to examine the thermodynamic properties of
this finite temperature system.
We have mentioned in the introduction that the methods of computing
the regularized action and thermodynamic quantities adopted in
References~\cite{Herzog:2008wg,Maldacena:2008wh} are not directly
applicable to non-planar black solutions.
We, therefore, use the subtraction method which
we describe the details here.
In what follows, ``the action'' is meant
to be the sum of the bulk action and the Gibbons-Hawking surface term.

We first analytically continue $x^+$ to $ix^+$ and
put the system into a box by the cutoff $r=r_B$.%
\footnote{
 The analytic continuation does not yield a Euclidean section,
 as noted by Herzog {\it et al.} \cite{Herzog:2008wg}.
 This is because though the system is not static, we do not
 have an appropriate ``rotation parameter'' whose simultaneous
 analytic continuation provides a Euclidean section as in
 Kerr black holes.
 We take the same stance as Herzog {\it et al.}: we carry out
 the analysis with the complex section because the on-shell
 action and other relevant quantities will all be real.
}
The cutoff $r_B$ is assumed to be much larger than the scale
$R$ but finite.
We subtract the action of the extremal
solution from the non-extremal one.
In doing so, it is instructed in Reference~\cite{Hawking:1982dh} to match
the metrics of those geometries at the wall $r=r_B$.%
\footnote{
 One does not attempt to match the second fundamental forms of
 the boundary surface, so it is too strong to say that we match
 the boundary ``geometry''.
}
In our case, we have a problem with the boundary matching:
since the $g_{--}$ component of the extremal metric (with $h=1$) in
Equation~(\ref{eq:k0nonExt}) is degenerate, there is no way we
can match the metrics at the wall.
As noted before, the $x^+$-constant box boundary slice 
has zero measure for the extremal case.
Therefore, we propose to match the boundary metric of the 
extremal geometry to the non-extremal one only for the $x^-$-constant
three dimensional slices parametrized by $x^{+,1,2}$.
We can achieve this by appropriately scaling those three
coordinates.
As for the $x^-$ direction, we scale this coordinate by a constant
and adjust the constant so that the resulting difference action is
finite in the limit $r_B\to\infty$.
It turns out that this constant is just unity for all the black solutions
discussed in this paper,
so we do not include the constant in the computations below.

Concretely, we have the analytically continued and
scaled extremal metric and the fields
\begin{align}\label{eq:k0scaled}
  ds^2 =& \bigg(\frac{r}{R}\bigg)^2
           \bigg[ \bigg(\frac{r}{R}\bigg)^2 H_B^2 dx^{+2}
        - 2i H_B dx^+dx^-
        + G_B^2 d\vec{x}^2 \bigg]+\bigg(\frac{R}{r}\bigg)^2 dr^2
  \;,\nonumber\\
  \Phi =& 0
  \;,\nonumber\\
  A =& i\bigg(\frac{r}{R}\bigg)^2 \frac{H_B}{R} dx^+
  \;,
\end{align}
where we have defined the scaling factors
\begin{align}
  H_B :=& \bigg( K(r_B)^{-2/3} \bigg\{\frac{h(r_B)-1}{(2bR)^2}
          + \bigg(\frac{r_B}{R}\bigg)^2 h(r_B) \bigg\} \bigg)^{1/2}
          \bigg(\frac{r_B}{R}\bigg)^{-1}
       	\;,\nonumber\\
  G_B :=& K(r_B)^{1/6}
  \;.
\end{align}
We then compute the action
\begin{equation}
  S_0 = S_{0\text{bulk}} + S_{0\text{GH}}
  \;,
\end{equation}
where $S_{0\text{bulk}}$ is the action (\ref{eq:theAction}) evaluated
on the solution (\ref{eq:k0scaled}). The second term $S_{0\text{GH}}$
is the Gibbons-Hawking term
\begin{equation}
  S_{0\text{GH}} = -\frac{1}{8\pi G_5} \int dx^4 \sqrt{g_B} (TrK_0)
  \;,
\end{equation}
where $g_B$ is the determinant of the boundary first fundamental form and
$(TrK_0)$ is the trace of the second fundamental form
with respect to the outward pointing unit normal vector.
Since the invariants in the action are not affected by the scaling,
the factors of $H_B$ and $G_B$ come in only from the metric
determinants.
Similarly, we compute the action $S = S_{\text{bulk}} + S_{\text{GH}}$
evaluated on the non-extremal solution (\ref{eq:k0nonExt}).
The difference $(S-S_0)$ is now finite in the limit $r_B\to\infty$ and
we have
\begin{equation}
  \lim_{r_B\to\infty}(S-S_0) = \frac{V_4}{16\pi G_5} \; \frac{r_H^4}{R^5}
  \;,\quad\text{with}\quad
  V_4 := \int dx^4
  \;.
\end{equation}
This result is in agreement with Reference~\cite{Herzog:2008wg} whose
authors have noticed that this is the same result as
the Schwarzschild-AdS$_5$ black brane.
In the discussion of the black hole solution, we will find that
the action is the same as the Schwarzschild-AdS$_5$ black hole.

\subsection{Thermodynamics}\label{subsec:BBThermo}

We now reproduce the thermodynamics of 
References~\cite{Herzog:2008wg,Maldacena:2008wh}
that follows from the regulated action.
For this purpose, we need to identify the temperature, chemical potential,
energy, charge and entropy of the system.
Let us start with the temperature.
We identify the temperature in the traditional way, that is,
we set $\beta = 2\pi/\kappa$ where $\kappa$ is the horizon
surface gravity.
In the computation of $\kappa$, the Killing vector field is
taken as
\begin{equation}\label{eq:KillingVF}
  \chi = \partial_+ + \Omega_H \partial_-
  \;,
\end{equation}
where the horizon angular velocity $\Omega_H$
is shown in Equation~(\ref{eq:OmegaH})
and it is measured in the units of the compact $x^-$ 
circumference in the coordinate length.
We then have
\begin{equation}
  \beta = \pi b R^3/r_H
  \;,
\end{equation}
and we set the (coordinate) circumference of $x^+$ to this value.
One can also obtain the same result by requiring the analytic
continuation of the metric (\ref{eq:k0nonExt}) to have the smooth
geometry in the $x^+$-$r$ slice.
[In doing so, one must bring the metric to the ADM form
(\ref{eq:ADMform}), just as in the Kerr black hole case.]

Let us now discuss the chemical potential of the system.
As discussed before,
our metric (\ref{eq:k0nonExt}) is stationary but not static and
it describes the black brane rotating in the compactified 
$x^-$ direction.
We then naturally interpret the angular momentum and velocity
as the charge and the conjugate chemical potential of the system,
respectively.
This offers
\begin{equation}\label{eq:chemp}
  \hat\mu := \Omega_H = \frac{1}{2 (bR)^2}
  \;,
\end{equation}
where the hat on $\mu$ reminds us that the angular velocity
is measured in the units of the $x^-$ circumference.
This identification, however, is subtle.
In the usual rotating black hole systems, a chemical potential
is taken to be the {\it difference} between the angular velocities
at the horizon and the boundary.
However, as we saw previously, our horizon angular velocity
is $\Omega_H$ and the boundary coordinate angular velocity diverges.
Hence the usual identification does not work.
Despite this subtlety,
we will see shortly that the definition (\ref{eq:chemp})
yields the consistent entropy, suggesting that the identification
is correct.

To obtain the energy, the charge (conjugate to the chemical potential
just defined) and the entropy,
we define the zero-loop saddle point free energy
\begin{align}\label{eq:k0free}
  F :=& - (16\pi G_5)V_3^{-1} \lim_{r_B\to\infty}(S-S_0)
     = - \beta (r_H^4/R^5)
     \nonumber\\
     =& - \frac{\pi^4 R^3}{4\beta^3\hat\mu^2}
  \;,
\end{align}
where $V_3$ represents the integration over $x^-$ and $\vec{x}$.
Notice that the product $G_5V_3^{-1}$ is dimensionless, hence so is $F$.
From this, we have
\begin{align}\label{eq:k0EQS}
  E =& \bigg( \frac{\partial F}{\partial\beta} \bigg)_{\hat\mu}
      - \hat\mu \beta^{-1} \bigg( \frac{\partial F}{\partial\hat\mu} \bigg)_\beta
    = r_H^4/R^5
  \nonumber\\
  Q =& - \beta^{-1} \bigg( \frac{\partial F}{\partial\hat\mu} \bigg)_\beta
    = - 4 b^2 r_H^4 / R^3
  \nonumber\\
  S =& \beta \bigg( \frac{\partial F}{\partial\beta} \bigg)_{\hat\mu} - F
    = 4 \pi b r_H^3 / R^2
  \;.
\end{align}
These results have been derived in
References~\cite{Herzog:2008wg,Maldacena:2008wh}.
These quantities, by construction, satisfy the first law.
One non-trivial result, though, is that the horizon area
[the three-volume of $x^+$-constant and $r=r_H$ slice of
the metric (\ref{eq:k0nonExt})] is exactly four times
the entropy derived above, apart from the overall factor
in the definition of the free energy (\ref{eq:k0free}).
This makes us confident of the chemical potential identified above.

To examine the local thermodynamic stability of the system, one can
compute the Hessian of $\beta(E-\hat\mu Q)-S$ with respect to
the thermodynamic variables $(r_H,b)$ and evaluate it at the on-shell
values of $(\beta,\hat\mu)$.
This gives $32\pi^2r_H^4/R^4$, which is always positive.
Therefore, the system is thermodynamically stable.

Finally, we note that the free energy (\ref{eq:k0free})
is negative, implying that the non-extremal black brane
solution is always thermodynamically preferred to
the thermal Schr\"odinger space without the black brane.
This means that the system does not possess the Hawking-Page
phase transition.

\section{Black Hole}\label{sec:BH}

It is common that a single gravitational
action supports the black solutions whose
horizon geometries are $\mathbb{R}^n$, $S^n$ and $\mathbb{H}^n$.
We find that it is also the case for the five dimensional
action (\ref{eq:theAction}).
In this section, 
we present the spherical solution
and the hyperbolic case is worked out in the appendix.
We will see that
the spherical case has a richer phase structure than the other solutions
such as the Hawking-Page phase transition.

\subsection{The Solution}\label{subsec:BHSoln}

Consider the Schwarzschild-AdS black hole solution of Type IIB
supergravity compactified on $S^5$
\begin{equation}
    ds^2 = \bigg(\frac{r}{R}\bigg)^2
            \big(-h\,dt^2 + R^2d\Omega_3^2\big)
         +\bigg(\frac{R}{r}\bigg)^2 h^{-1}dr^2 + R^2d\Omega_5^2
  \;,
\end{equation}
where $h:=1+(R/r)^2-(r_0R/r^2)^2$ and $r_0$ is the non-extremality
parameter in that the extremal solution is given by $r_0=0$.
This parameter can be re-expressed in terms of $r_H$ by solving
$h(r_H)=0$.
The metric of the three sphere $d\Omega_3^2$
can be written as
\begin{equation}
  R^2d\Omega_3^2 = \eta^2 + dX^2
  \;,
\end{equation}
where we have defined
\begin{equation}
  \eta := \frac{R}{2} (d\psi+\cos\theta d\phi)
  \;,\quad\text{and}\quad
  dX^2 := \bigg(\frac{R}{2}\bigg)^2(d\theta^2 + \sin^2\theta d\phi^2)
  \;.
\end{equation}
Then the metric above can be trivially rewritten as
\begin{equation}
    ds^2 = \bigg(\frac{r}{R}\bigg)^2
            \big(-h\,dt^2 + \eta^2 + dX^2 \big)
         +\bigg(\frac{R}{r}\bigg)^2 h^{-1}dr^2 + R^2d\Omega_5^2
  \;.
\end{equation}
One surely notices the similarity of this metric to the metric
(\ref{eq:1stMetric}).

Guided by this analogy, consider the following set of a metric
and fields which is very similar to the truncated version of
Equations (\ref{eq:k0tybasis}),
\begin{align}
  ds^2 =& K^{-2/3}\bigg(\frac{r}{R}\bigg)^2 
          \bigg[-(1+b^2r^2)\,h\, dt^2 -2b^2r^2 h\, dt\, \eta
          +(1-b^2r^2h) \eta^2 + K dX^2 \bigg]
          \nonumber\\
          &+ K^{1/3}\bigg(\frac{R}{r}\bigg)^2 h^{-1}dr^2
  \;,\nonumber\\
  \Phi =& -\frac{1}{2}\ln K
  \;,\nonumber\\
  A =& K^{-1}\bigg(\frac{r}{R}\bigg)^2 b\, (\,h\, dt + \eta)
  \;,
\end{align}
where $K:=1-(h-1)b^2r^2$.
One can check that this set indeed is a solution to the equations
of motion that follow from the five dimensional
action (\ref{eq:theAction}).%
\footnote{
 I thank M. Hanada for double checking this solution.
}
Notice that this solution reduces to the Schwarzschild-AdS black hole
solution by setting $b=0$, like the KK-reduced solution
(\ref{eq:k0tybasis}) becomes the Schwarzschild-AdS black brane.
Also, just as in the AdS case, this solution's isometry group
is the maximal compact subgroup of the Schr\"odinger group.
Thus, we see that this is the (asymptotically)
Schr\"odinger space which has $S^2$-foliation for the slice of
constant $t$ and $\eta$.
To identify the Hamiltonian generator and the compactified
direction, we need to define the light-cone-like
basis and we introduce the following sets of the coframe
\begin{align}\label{eq:coframe}
  \omega^+ = bR(dt+\eta)
  \;,\qquad
  \omega^- = \frac{1}{2bR}(dt-\eta)
  \;,
  \nonumber\\
  \omega^\theta = \frac{R}{2} d\theta
  \;,\qquad
  \omega^\phi = \frac{R}{2} \sin\theta d\phi
  \;,\qquad
  \omega^r = dr
  \;,
\end{align}
and the frame
\begin{align}\label{eq:frame}
  e_+ = \frac{1}{2bR}(\partial_t + \frac{2}{R}\partial_\psi)
  \;,\qquad
  e_- = bR(\partial_t - \frac{2}{R}\partial_\psi)
  \;,
  \nonumber\\
  e_\theta = \frac{2}{R} \partial_\theta
  \;,\qquad
  e_\phi = \frac{2}{R} \big( -\cot\theta\partial_\psi
        + \frac{1}{\sin\theta}\partial_\phi \big)
  \;,\qquad
  e_r = \partial_r
  \;.
\end{align}
Note that we have $\omega^i e_j = \delta^i_j$, as required
for a frame and the corresponding coframe and also observe
that $\omega^\pm$ are defined similar to the light-cone
coordinates of Equations~(\ref{eq:LC}).
This clearly is a non-coordinate basis and we have the non-vanishing
Lie product
\begin{equation}\label{eq:Lie}
  [e_\theta , e_\phi] = 2b\,e_+ - \frac{1}{bR^2}e_-
  	-\frac{2}{R}\cot\theta\, e_\phi
  \;.
\end{equation}
Because of this non-vanishing Lie bracket, the usual definitions
of geometric quantities based on the coordinate basis must be
modified (see MTW~\cite{MTW}).
Among other things, the Christoffel symbols $\Gamma^\alpha_{\mu\nu}$
defined with respect to a coordinate basis
are modified and we no longer necessarily
have the symmetry $\Gamma^\alpha_{\mu\nu}=\Gamma^\alpha_{\nu\mu}$
in the non-coordinate basis.

Given the new coframe, we can now write the solution as
\begin{align}\label{eq:k1nonExt}
  ds^2 =& K^{-2/3} \bigg(\frac{r}{R}\bigg)^2
          \bigg[ -\bigg\{\frac{h-1}{(2bR)^2}
          + \bigg(\frac{r}{R}\bigg)^2 h \bigg\} \omega^{+2}
          - (1+h) \omega^+\omega^-
          + (bR)^2(1-h)\omega^{-2}
        \nonumber\\
        &+ K \vec\omega^2 \bigg]
         + K^{1/3} \bigg(\frac{R}{r}\bigg)^2 h^{-1}\omega^{r2}
  \;,\nonumber\\
  \Phi =& -\frac{1}{2}\ln K
  \;,\nonumber\\
  A =& K^{-1}\bigg(\frac{r}{R}\bigg)^2 b \,
     \bigg\{ \frac{h+1}{2bR} \omega^+ + bR(h-1) \omega^- \bigg\}
  \;,
\end{align}
where we have defined $\vec\omega := (\omega^\theta,\omega^\phi)$.
Here, one sees that the coframe is chosen so that the solution
appears congruous to the black brane solution (\ref{eq:k0nonExt}).
In parallel to the black brane case, we interpret
$\omega^+$ as the direction of time and $\omega^-$ as the
compactified direction with some circumference.
Though obvious, it is an interesting exercise to check that
the equations of motion are satisfied in the new frame and coframe
with the modified geometric quantities.%
\footnote{
 Some care is necessary for this.
 For example, we have $F_{\mu\nu}:=A_{\nu;\mu}-A_{\mu;\nu}$ and
 we may not replace the semi-colons to just commas, because
 the symmetry of the Christoffel symbols is lost.
 Another subtle point is that the definition of Riemann tensor
 involves the non-trivial derivatives with respect to $e_\theta$
 which is different from $\partial_\theta$ by the factor of
 $2/R$ and this must be done with care.
}

We remark that for this solution to make sense, we must impose
the restriction
\begin{equation}\label{eq:restriction}
  bR < 1
  \;.
\end{equation}
This is because we have $K=1-(bR)^2+(bRr_0/r)^2$,
and without the restriction, $K$ can become zero or negative.
This is a special condition for the black hole solution and
we do not have it for the black brane case.

As in the previous section, the geometry (\ref{eq:k1nonExt})
has the ill-defined causal structure.
We can bring the metric to the ADM form
identical to Equation~(\ref{eq:ADMform}) with the obvious
replacements.
A quick inspection reveals that with the function $h$ of
the black hole solution, the ADM form is highly pathological.
For example, the extremal case ($r_0=0$) with the condition
(\ref{eq:restriction}) has negative definite lapse function
squared.
Therefore, as in the black brane case, we must give up on
the causal development of the hypersurface defined by constant
$\omega^+$.
Currently the consequences of this observation are unclear
and it must be investigated in future
whether this has something to do with the holographic dual of
{\it non-relativistic} CFT on $S^2$.

\subsection{The Difference Action}\label{subsec:BHDiffAct}

Let us now proceed to compute the on-shell difference action following
the procedure described in the previous section.
Unlike the previous case, the $\omega^-$ direction is in general
not degenerate
for the extremal ($r_0=0$) nor non-extremal ($r_0\neq0$)
black hole solutions.
However, simple scaling of the cobasis cannot achieve the full matching
of the extremal $r=r_B$ boundary metric to the non-extremal one.
Therefore, we match the metrics only for the
$\omega^-$-constant three-slices parametrized by
$\omega^{+,\theta,\phi}$, just as was done in the previous section.

We analytically continue $\omega^+$ to $i\omega^+$ and set the cutoff
at $r=r_B$.
We then have the scaled extremal metric and fields
\begin{align}\label{eq:k1scaled}
  ds^2 =& K_0^{-2/3} \bigg(\frac{r}{R}\bigg)^2
          \bigg[ -\bigg\{\frac{h_0-1}{(2bR)^2}
          + \bigg(\frac{r}{R}\bigg)^2 h_0 \bigg\} H_B^2\,\omega^{+2}
          - (1+h_0) H_B\, \omega^+\omega^-
  \nonumber\\
        &+ (bR)^2(1-h_0)\omega^{-2}
         + K_0 G_B^2\,\vec\omega^2 \bigg]
         + K_0^{1/3} \bigg(\frac{R}{r}\bigg)^2 h_0^{-1}\omega^{r2}
  \;,\nonumber\\
  \Phi =& -\frac{1}{2}\ln K_0
  \;,\nonumber\\
  A =& K_0^{-1}\bigg(\frac{r}{R}\bigg)^2 b \,
     \bigg\{ \frac{h_0+1}{2bR} H_B\, \omega^+ + bR(h_0-1) \omega^- \bigg\}
  \;,
\end{align}
where we have defined
\begin{align}
  H_B :=& \bigg( K(r_B)^{-2/3} \bigg\{\frac{h(r_B)-1}{(2bR)^2}
          + \bigg(\frac{r_B}{R}\bigg)^2 h(r_B) \bigg\} \bigg)^{1/2}
  \nonumber\\
         &\times \bigg( K_0(r_B)^{-2/3} \bigg\{\frac{h_0(r_B)-1}{(2bR)^2}
          + \bigg(\frac{r_B}{R}\bigg)^2 h_0(r_B) \bigg\} \bigg)^{-1/2}
  \;,\nonumber\\
  G_B :=& \{ K(r_B)/K_0(r_B) \}^{1/6}
  \;,\nonumber\\
  h_0 :=& 1 + (R/r)^2
  \;,\nonumber\\
  K_0 :=& 1 - (bR)^2
  \;.
\end{align}
Next we compute the action
\begin{equation}
  S_0 = S_{0\text{bulk}} + S_{0\text{GH}}
  \;,
\end{equation}
where $S_{0\text{bulk}}$ is the action (\ref{eq:theAction}) evaluated
on the solution (\ref{eq:k1scaled}) and the second term $S_{0\text{GH}}$
is the Gibbons-Hawking term.
Notice that due to the scaling in (\ref{eq:k1scaled}),
the coframe (\ref{eq:coframe}) and frame (\ref{eq:frame}) must be
scaled accordingly, including the factor of ``$\,i\,$'' that comes from
the analytic continuation.
This results in the modified Lie bracket (\ref{eq:Lie})
which in turn affects the non-coordinate based geometric quantities.
Using those quantities, one can check that the invariants in
the action $S_0$ are independent of the scaling and the scaling
factors come in only from the metric determinants, as expected.

Similarly, we compute the action $S = S_{\text{bulk}} + S_{\text{GH}}$
evaluated on the non-extremal solution (\ref{eq:k1nonExt}).
The difference $(S-S_0)$ is finite in the limit $r_B\to\infty$ and
we have
\begin{equation}\label{eq:k1action}
  \lim_{r_B\to\infty}(S-S_0) = \frac{V_4}{16\pi G_5}
  (r_H^2 - R^2)\frac{r_H^2}{R^5}
  \;,\quad\text{with}\quad
  V_4 := \int \omega^+\wedge\omega^-\wedge\omega^\theta\wedge\omega^\phi
  \;.
\end{equation}
Rather surprisingly, this action is the same as
the Schwarzschild-AdS$_5$ black hole.

\subsection{Thermodynamics}\label{subsec:BHThermo}

In this subsection, we examine the thermodynamics following from
the action calculated in Equation~(\ref{eq:k1action}).
We proceed in parallel with the previous section.

First, we identify the temperature.
We can compute this just as in the previous section, either through
the surface gravity [using the same Killing vector field
as Equation~(\ref{eq:KillingVF}) with the replacements
$\partial_\pm \to e_\pm$] or by requiring the smooth geometry.
The result is
\begin{equation}\label{eq:k1beta}
  \beta = 2\pi b r_H R^3 / (2r_H^2 + R^2)
  \;.
\end{equation}
Now, it is a little naive to identify this to the ``circumference''
of $\omega^+$, because it is not a coordinate basis.
However observe that at each fixed point on $S^2$, parametrized by
($\theta,\phi$), $\omega^+$ {\it is} a coordinate basis.
We thus write $V_4$ of Equation~(\ref{eq:k1action}) as
\begin{equation}
  V_4 = \beta\, V_3
  \;,
\end{equation}
where $V_3$ includes the ``circumference'' of the $\omega^-$ direction.

Second, we identify the chemical potential to the angular velocity
of the horizon as before and it turns out that it is the same
\begin{equation}\label{eq:k1chemp}
  \hat\mu := \Omega_H = \frac{1}{2 (bR)^2}
  \;.
\end{equation}

To obtain the thermodynamic quantities, we define the
zero-loop saddle point free energy
\begin{align}\label{eq:k1free}
  F :=& - (16\pi G_5)V_3^{-1} \lim_{r_B\to\infty}(S-S_0)
     = - \beta (r_H^2 - R^2)\frac{r_H^2}{R^5}
     \nonumber\\
     =& - \frac{1}{16R\beta^3\hat\mu^2}
       \big[ f(\beta,\hat\mu)-2\beta^2\hat\mu \big]
       \big[ f(\beta,\hat\mu)-6\beta^2\hat\mu \big]
  \;,
\end{align}
where $f:=(\pi R)^2(1+\sqrt{1-(4\beta^2\hat\mu)/(\pi R)^2})$
and this is defined purely for the presentation purpose and
we do not mean anything deeper in this function $f$.
Then we compute the thermodynamic quantities as in 
Equations~(\ref{eq:k0EQS}) and obtain
\begin{equation}\label{eq:k1EQS}
  E = (r_H^2+2R^2)\frac{r_H^2}{R^5}
  \;,\qquad
  Q = - \frac{2 b^2 r_H^2}{R^3} (2r_H^2+R^2)
  \;,\qquad
  S = 4\pi \frac{b r_H^3}{R^2}
  \;.
\end{equation}
Again, by construction, these quantities satisfy the first law.
The entropy, thus calculated, is consistent with the quarter-area
law which is a non-trivial result and indicates the right
identification of the chemical potential.

The local thermodynamic stability can be examined by
computing the Hessian of $\beta(E-\hat\mu Q)-S$ with respect to
the thermodynamic variables $(r_H,b)$ and evaluate it at the on-shell
values of $(\beta,\hat\mu)$.
The result is
\begin{equation}
  \text{Hessian} =
    \frac{64\pi^2r_H^4(2r_H^2-R^2)(r_H^2+R^2)}{R^4(2r_H^2+R^2)^2}
  \;.
\end{equation}
This implies the stability threshold $r_H = R/\sqrt{2}$ which
translates to the following curve in the $\hat\mu$-$T$ parameter
space:
\begin{equation}\label{eq:instability}
  \hat\mu = \frac{\pi^2}{4} R^2 T^2
  \;.
\end{equation}
The mechanism of this instability is very similar to the usual
Schwarzschild-AdS black hole, namely, it is the merger point
of the large and small black holes in the parameter space.
Though we have not explicitly discussed the small black hole,
it exists.
A quick way to see this is to note that there are two solutions
to Equation~(\ref{eq:k1beta}) with respect to the variable $r_H$
and these solutions degenerate at the threshold.

Let us now discuss the Hawking-Page phase transition \cite{Hawking:1982dh}.
It is clear from Equation~(\ref{eq:k1free}) that the free energy
changes sign at $r_H=R$.
This implies that the thermodynamically preferred solution changes
from the thermal space without black hole to the space with black
hole, and {\it vice versa}, at the threshold.
This is the Hawking-Page phase transition.
The threshold $r_H=R$ can be expressed in the parameter space
of $\hat\mu$ and $T$ as
\begin{equation}\label{eq:HPtrans}
  \hat\mu = \frac{2\pi^2}{9} R^2 T^2
  \;,
\end{equation}
and the higher temperature side of this curve is the black hole
phase.
Comparing Equations~(\ref{eq:instability}) and (\ref{eq:HPtrans}),
we see that the black hole instability line is inside
the thermal Schr\"odinger geometry phase.

Finally, we note that the restriction (\ref{eq:restriction}) requires
\begin{equation}\label{eq:muRestrict}
  \hat\mu > \frac{1}{2}
  \;,\quad\text{and}\quad
  T > \frac{\sqrt{2}}{\pi R}
  \;,
\end{equation}
where the critical temperature coincides with
the crossing point of the $\hat\mu=1/2$ line
and the local thermodynamic instability curve (\ref{eq:instability}).
Recall that the chemical potential $\hat\mu$ is measured in the units
of the circumference of the compactified direction.
Therefore, this restriction may be alternatively 
interpreted as the minimal size of the compactification.
\\

\begin{figure}[h]
{
\centerline{\scalebox{0.5}{\includegraphics{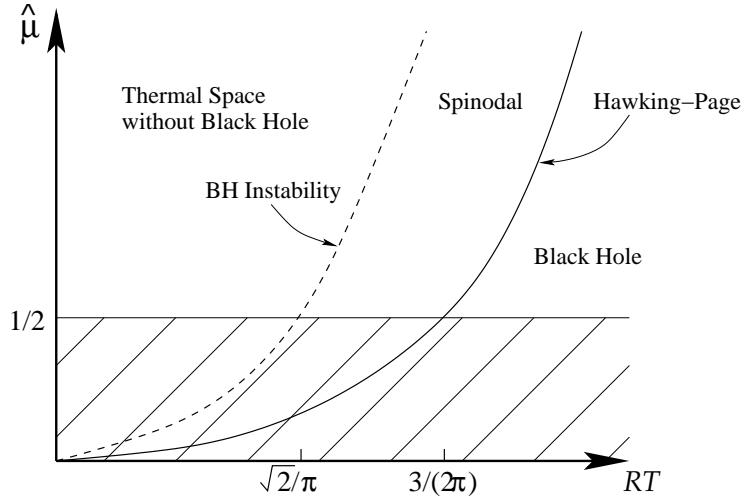}}}
\caption{\footnotesize
  The phase diagram in the parameter space of the chemical
  potential $\hat\mu$ and the temperature measured in the
  units of $R$.
  The dotted and solid curves represent the instability
  curve (\ref{eq:instability}) and the Hawking-Page phase
  transition curve (\ref{eq:HPtrans}), respectively.
  The line at $\hat\mu=1/2$ is the restriction (\ref{eq:muRestrict}).
  The area between the dotted and solid curves is the spinodal
  phase where the locally stable black hole solution exists
  but it is not energetically preferred.
}\label{fig:SchrPhase}%
}
\end{figure}

The discussions above are summarized in the phase diagram
Figure~\ref{fig:SchrPhase}.
%
%

\section{Discussions}\label{sec:discussions}

In this paper, we have discussed the thermodynamics of the black
solutions to the action that describes the space with
Schr\"odinger symmetry.
The black hole and black hyperboloid solutions are newly found
solutions and also we have proposed a way to compute the finite
actions.
For the black hole case, we have found the Hawking-Page
phase transition, as well as the instability lines in the phase
diagram.

We have shown that the on-shell actions of all solutions
discussed in this paper are identical to those of
the black solutions in AdS.
This seems to imply that there is a symmetry that transforms
the asymptotically Schr\"odinger solutions to the corresponding
asymptotically AdS solutions.
For the planar case, Reference~\cite{Maldacena:2008wh} explicitly
show such a symmetry.
Therefore, it is likely that the symmetry transformation of
the reference can be generalized to the non-planar solutions
discussed in this paper.
If this expectation is borne out,
the method similar to the one proposed in 
Appendix C.2 of Reference~\cite{Maldacena:2008wh} can be
applied to the solutions other than the black branes
and the resulting thermodynamic quantities
computed this way would agree with our results in this paper.

It is undeniable that the matching procedure proposed in
this paper is highly {\it ad hoc}.
Though this procedure produces the reasonable results,
further understanding of this is required.
This issue is closely related to the 
(so far not so clear) physical meaning of
the $x^-$ coordinate, because the procedure gives up on
the matching of the boundary first fundamental forms for
this component.
Perhaps, this is related to the fact that the boundary
CFT is defined on the coordinate that is not including $x^-$.

We have mentioned in Section~\ref{subsec:BBThermo} that there is
a certain subtlety in identifying the chemical potential, because
the difference of the angular velocities at the horizon and the
boundary diverges.
However, we have shown that our identification of the chemical
potential yields the entropy that is consistent with
the quarter-area law.
On the contrary,
it is advocated in References~\cite{Son:2008ye,Adams:2008wt}
that the chemical potential should be identified with the coefficient
of $1/r^2$ term in the $g_{++}$ component.
This identification yields the entropy that is
{\it not} consistent with the quarter-area law.
However, it is not clear to us what is inappropriate with
their identification.
Clarification on this issue is desired.

As pointed out in the ends of Sections~\ref{subsec:BBSoln} and
\ref{subsec:BHSoln}, the spacetimes discussed in this paper
are causally pathological.
The problem is clear when we bring the metric into the ADM form
with respect to the $x^+$ direction which the Galilean holography
seems to uniquely selects as the time.
It is likely that this behavior of the spacetime is closely related
to the fact that this is the (conjectured) dual of non-relativistic
conformal field theories.
We expect that investigation into this issue would deepen our
understanding of the holography.

It is known that the R-charged black holes possess a rich phase
structure \cite{Cvetic:1999ne,Yamada:2007gb}.
It is possible to include the R-charge into the Galilean holography
by applying the Null Melvin Twist to the spinning D3-brane
geometry
\cite{Kraus:1998hv,Russo:1998by,Cvetic:1999xp} and taking the near
horizon limit of the resulting geometry.
(The RR potential, in this case, becomes non-trivial and this itself
is interesting to investigate.)
From this solution, one could obtain the five dimensional action
and its spherical (and hyperbolic) solution.
We then able to observe the interplay between the chemical potential
discussed in this paper and the one for the R-symmetry.
It would be very interesting to see how the phase structure of
the R-charged black holes would be extended.

In this paper, we have exclusively worked out the gravity side
of the proposed Galilean holography.
It is of natural interest to investigate into
the CFT side of the corresponding
phenomena found in this work.
Most importantly, we would like to see in detail how
the non-relativistic conformal field theory behaves in accordance
with the Hawking-Page phase transition found in this paper.
Learning from the AdS/CFT correspondence, it is likely that
the phase transition corresponds to the confinement/deconfinement
phase transition in the CFT side.
It would be intriguing to see how this actually works
in the non-relativistic set up.

Meanwhile, we should always keep in mind that this holography
is initiated in hope to understand the real world strongly coupled
non-relativistic field theories.
The gravity duals being discussed in the literature (including this work)
most likely do not have the real world field theory duals.
But as in the AdS/CFT correspondence, we expect the Galilean holography
to provide important universal properties of non-relativistic field
theories.
The fact is that the experimental data exists on the field theory side
of the proposed duality.
It is crucial to see if one could come up with the gravity dual
that is consistent with the experimental data.

\bigskip
\bigskip

\section*{Acknowledgments}

I would like to thank Andrei Kryjevski for drawing my attention
to Son's original work \cite{Son:2008ye}.
I benefited from the discussion session held at
Weizmann Institute, especially from
Ofer Aharony and Zohar Komargodski.
I would also like to thank Umpei Miyamoto for valuable discussions.
I thank
A.~Kryjevski and M.~Hanada for the partial involvements in this project.
This work was supported by the Lady Davis fellowship.

\bigskip
\bigskip

\appendix

\section{Black Hyperboloid}\label{app}

In this appendix, we present the black hyperboloid solution
and examine its thermodynamics.
The discussion follows the structure of the main text.

We would like to find the solution to the action
(\ref{eq:theAction}) with $\mathbb{H}^2$ horizon.
For this purpose, we find the appropriate form of the metric for
$\mathbb{H}^3$ as
\begin{equation}
  ds^2 = \eta^2 + dX^2
  \;,
\end{equation}
with
\begin{equation}
  \eta = \frac{R}{2}(d\psi + \cosh\chi d\phi)
  \;,\quad\text{and}\quad
  dX^2 = \bigg(\frac{R}{2}\bigg)^2
        (d\chi^2-\sinh^2\chi d\phi^2)
  \;.
\end{equation}
We then find the desired solution
\begin{align}
  ds^2 =& K^{-2/3} \bigg(\frac{r}{R}\bigg)^2
          \bigg[ -\bigg\{\frac{h-1}{(2bR)^2}
          + \bigg(\frac{r}{R}\bigg)^2 h \bigg\} \omega^{+2}
          - (1+h) \omega^+\omega^-
          + (bR)^2(1-h)\omega^{-2}
        \nonumber\\
        &+ K (\omega^{\chi 2}-\omega^{\phi 2}) \bigg]
         + K^{1/3} \bigg(\frac{R}{r}\bigg)^2 h^{-1}\omega^{r2}
  \;,\nonumber\\
  \Phi =& -\frac{1}{2}\ln K
  \;,\nonumber\\
  A =& K^{-1}\bigg(\frac{r}{R}\bigg)^2 b \,
     \bigg\{ \frac{h+1}{2bR} \omega^+ + bR(h-1) \omega^- \bigg\}
  \;,\nonumber\\
  h :=& 1 - \bigg(\frac{R}{r}\bigg)^2 - \bigg(\frac{r_0R}{r^2}\bigg)^2
  \;,\nonumber\\
  K :=& 1-(h-1)b^2r^2
        = 1+(bR)^2 + \bigg(\frac{r_0bR}{r}\bigg)^2
  \;,
\end{align}
where we have defined the coframe
\begin{align}
  \omega^+ = bR(dt+\eta)
  \;,\qquad
  \omega^- = \frac{1}{2bR}(dt-\eta)
  \;,
  \nonumber\\
  \omega^\chi = \frac{R}{2} d\chi
  \;,\qquad
  \omega^\phi = \frac{R}{2} \sinh\chi d\phi
  \;,\qquad
  \omega^r = dr
  \;,
\end{align}
and the frame
\begin{align}
  e_+ = \frac{1}{2bR}(\partial_t + \frac{2}{R}\partial_\psi)
  \;,\qquad
  e_- = bR(\partial_t - \frac{2}{R}\partial_\psi)
  \;,
  \nonumber\\
  e_\chi = \frac{2}{R} \partial_\chi
  \;,\qquad
  e_\phi = \frac{2}{R} \big( -\coth\chi\partial_\psi
        + \frac{1}{\sinh\chi}\partial_\phi \big)
  \;,\qquad
  e_r = \partial_r
  \;.
\end{align}
With this non-coordinate basis, the non-vanishing Lie bracket is
\begin{equation}
  [e_\chi , e_\phi] = -2b\,e_+ + \frac{1}{bR^2}e_-
  	-\frac{2}{R}\coth\chi\, e_\phi
  \;.
\end{equation}
Notice that the non-extremality parameter $r_0$ is expressed as
\begin{equation}
  r_0 = \frac{r_H^2}{R}\sqrt{1-(R/r_H)^2}
  \;,
\end{equation}
so we have the extremality at $r_H = R$ and the horizon radius must
be larger than this value.

One can proceed to compute the difference action just as in the main
text and find that
\begin{equation}
  \lim_{r_B\to\infty}(S-S_0) = \frac{V_4}{16\pi G_5}
  (r_H^2 + R^2)\frac{r_H^2}{R^5}
  \;,\quad\text{with}\quad
  V_4 := \int \omega^+\wedge\omega^-\wedge\omega^\chi\wedge\omega^\phi
  \;.
\end{equation}
This is the same as the Schwarzschild-AdS$_5$ black hyperboloid.
The temperature can be computed in the usual way and one obtains
\begin{equation}
    \beta = 2\pi b r_H R^3 / (2r_H^2 - R^2)
  \;.
\end{equation}
Since we have the restriction $r_H>R$, we see that this solution does
not have zero temperature limit.
The chemical potential is the same as other cases:
\begin{equation}
  \hat\mu = \frac{1}{2(bR)^2}
  \;.
\end{equation}

To obtain the thermodynamic quantities, we define the
zero-loop saddle point free energy
\begin{align}
  F :=& - (16\pi G_5)V_3^{-1} \lim_{r_B\to\infty}(S-S_0)
     = - \beta (r_H^2 + R^2)\frac{r_H^2}{R^5}
     \nonumber\\
     =& - \frac{1}{32R^3\beta^3\hat\mu^2}
       f(\beta,\hat\mu)^2
       \big[ f(\beta,\hat\mu)+6\beta^2\hat\mu \big]
  \;,
\end{align}
where $f:= \pi R^2(1+\sqrt{1+(4\beta^2\hat\mu)/(\pi R)^2})$.
Then we compute the thermodynamic quantities as in 
Equations~(\ref{eq:k0EQS}) and obtain
\begin{equation}
  E = (r_H^2-2R^2)\frac{r_H^2}{R^5}
  \;,\qquad
  Q = - \frac{2 b^2 r_H^2}{R^3} (2r_H^2-R^2)
  \;,\qquad
  S = 4\pi \frac{b r_H^3}{R^2}
  \;.
\end{equation}
By construction, these quantities satisfy the first law and
the entropy is consistent with the quarter-area law.

The local thermodynamic stability can be examined by
computing the Hessian of $\beta(E-\hat\mu Q)-S$ with respect to
the thermodynamic variables $(r_H,b)$ and evaluate it at the on-shell
values of $(\beta,\hat\mu)$.
The result is
\begin{equation}
  \text{Hessian} =
    \frac{64\pi^2r_H^4(2r_H^2+R^2)(r_H^2-R^2)}{R^4(2r_H^2-R^2)^2}
  \;.
\end{equation}
Because of the restriction $r_H>R$, we see that the system is
always locally stable.
Notice that for the range of the horizon radius
$R<r_H<\sqrt{2}R$, we have $E<0$, but the system is stable.

Finally we remark that the free energy is always negative,
implying that the black solution is always preferred and
there is no Hawking-Page phase transition.

\pagebreak


\end{document}